\newcounter{myctr}
\def\myitem{\refstepcounter{myctr}\bibfont\noindent\ifnum\themyctr>9\else\phantom{0}\fi\hangindent17pt\themyctr.\enskip}

\documentclass{ws-ijqi}
\usepackage[numbers,sort&compress]{natbib}

\usepackage{bbm}
\usepackage{hyperref}
\usepackage{xcolor}

\let\originalleft\left
\let\originalright\right
\renewcommand{\left}{\mathopen{}\mathclose\bgroup\originalleft}
\renewcommand{\right}{\aftergroup\egroup\originalright}
\renewcommand{\right}{\aftergroup\egroup\originalright}

\def\Tr{\hbox{Tr}}

\newcommand{\id}{\mathbbm{1}}

\newcommand{\qunr}[1]{ \widetilde{\qfi}_{\mathsf{unr}, #1}}

\newcommand{\qfi}{\mathcal{Q}}

\usepackage[textsize=footnotesize]{todonotes}

\begin{document}

%%%%%%%%%%%%%%%%%%%%% Publisher's Area please ignore %%%%%%%%%%%%%%
\catchline{}{}{}{}{}
%%%%%%%%%%%%%%%%%%%%%%%%%%%%%%%%%%%%%%%%%%%%%%%%%%%%%%%%%%%%%%%%%%%

\title{Quantum frequency estimation with conditional states of continuously monitored independent dephasing channels}

\author{Francesco Albarelli}

\address{Department of Physics, University of Warwick\\
Coventry CV4 7AL, United Kingdom\\
francesco.albarelli@gmail.com}

\author{Matteo A. C. Rossi}

\address{QTF Centre of Excellence, Turku Centre for Quantum Physics\\
Department of Physics and Astronomy, University of Turku\\
FI-20014 Turun Yliopisto, Finland \\
matteo.rossi@utu.fi}

\author{Marco G. Genoni}

\address{Quantum Technology Lab, Dipartimento di Fisica ``Aldo Pontremoli''\\
Università degli Studi di Milano\\
IT-20133, Milan, Italy \\
marco.genoni@fisica.unimi.it}

\maketitle

% \begin{history}
% \received{Day Month Year}
% \revised{Day Month Year}
% %\accepted{Day Month Year}
% %\comby{(xxxxxxxxxx)}
% \end{history}

\begin{abstract}
We discuss the problem of estimating a frequency via $N$-qubit probes undergoing independent dephasing channels that can be continuously monitored via homodyne or photo-detection.
We derive the corresponding analytical solutions for the conditional states, for generic initial states and for arbitrary efficiency of the continuous monitoring.
For the detection strategies considered, we show that: i) in the case of perfect continuous detection, the quantum Fisher information (QFI) of the conditional states is equal to the one obtained in the noiseless dynamics; ii) for smaller detection efficiencies, the QFI of the conditional state is equal to the QFI of a state undergoing the (unconditional) dephasing dynamics, but with an effectively reduced noise parameter.
% Besides the relevance for quantum metrology, our results describe from a different perspective that pure dephasing can in fact be described by the action of a classical stochastic fields entering in the system Hamiltonian.
\end{abstract}


\section{Introduction}

Peculiar properties of quantum mechanics, such as entanglement and squeezing, can be exploited to design quantum estimation schemes with a precision that cannot be obtained via purely classical resources~\cite{Caves1981,Braunstein1994,GiovannettiNatPhot,QuantumSensingRMP,Braun2018,Pirandola2018}.
Frequency estimation is one of the most paradigmatic examples of quantum-enhanced metrology: the frequency $\omega$ is unitarily encoded in the state of $N$ qubits as a rotation around one of the axes of each qubit's Bloch sphere.
If the qubits are initially prepared in an entangled state, the precision on the estimation of the parameter $\omega$ follows the so-called Heisenberg scaling  $1/N$, yielding thus an enhancement of order $O(\sqrt{N})$ compared to the \emph{classical} SQL scaling $1/\sqrt{N}$ that is obtained when the qubits are prepared in a separable state~\cite{Bollinger1996}.
For frequency estimation with an open system, where each qubit undergoes the same dephasing dynamics, it is known that the Heisenberg scaling is readily lost for non-zero noise: initial quantum correlations in the probe state can only provide a constant enhancement~\cite{Huelga97,Haase2018}.
This is in fact a specific instance of a more general no-go result that applies to generic noisy quantum metrology protocols~\cite{EscherNatPhys,KolodynskyNatComm}.

In this manuscript, we consider a dephasing noise caused by the interaction of the qubits with independent quantum environments and not by classical fluctuations in the Hamiltonian.
We assume that all these environmental degrees of freedom are (at least partially) accessible and measured continuously in time, leading to a stochastic conditional dynamics for the $N$-qubit state.
Continuous measurements have been proposed for different quantum estimation problems, e.g. for magnetometry~\cite{Geremia2003,Molmer2004,Madsen2004,Stockton2004,Albarelli2017a}, phase tracking~\cite{Yonezawa2012,Shankar2019}, waveform estimation~\cite{Tsang2011}, state estimation and generic dynamical parameters~\cite{Gambetta2001,Ralph2011a,Gammelmark2013a,Gammelmark2014,KiilerichPC,KiilerichHomodyne,Six2015,Calypso,Genoni2017,Cortez2017,Ralph2017}.
In particular, in our previous paper~\cite{Albarelli2018a} we have shown the usefulness of continuous monitoring to counteract the effect of noise in frequency estimation.

In~\cite{Albarelli2018a} we have shown that, for initial GHZ states and perfectly efficient continuous monitoring of the environment, the Heisenberg scaling can in fact be restored, both for dephasing and for noise transversal to the Hamiltonian generating the rotation.
While several other approaches around the no-go theorems in the presence of pure dephasing, obtaining super-classical scalings by exploiting temporal correlations of the environment, were already presented in the literature~\cite{Chin12,Smirne2015a,Macieszczak2015,Haase2017}, this was the first example where the effect of Markovian dephasing, caused by the interaction with a quantum environment, has been counteracted.

In this work we focus on Markovian dephasing noise only and we expand on the results of~\cite{Albarelli2018a}, presenting considerably more general results, that are valid for arbitrary initial states.
Furthermore, we now present a fully analytical solution of the dynamics for generic efficiency of the monitoring, for both continuous photo-detection and homodyne detection, while previously the case of inefficient detection was treated numerically.

The manuscript is structured as follows: in Sec.~\ref{s:conditional} we give a brief introduction on the dynamics of $N$ qubits rotating with frequency $\omega$ and subjected to dephasing noise; in particular we discuss both the unconditional and the conditional dynamics due to either continuous photo-detection and homodyne detection.
In Sec.~\ref{s:estcontinuous} we review the methods needed to assess quantum parameter estimation in continuously monitored quantum systems, while in Sec.~\ref{s:results} we present our results, that is, the analytical solutions for the conditional states under continuous monitoring and the consequences on the effectiveness of these strategies for frequency estimation.
  % in the presence of dephasing.
 Sec.~\ref{s:conclusions} concludes the manuscript with some remarks and outlooks.

\section{Conditional and unconditional dynamics}\label{s:conditional}
We consider a system of $N$ qubits (e.g. two-level atoms) undergoing a collective rotation with frequency $\omega$ around the $z$ axis and with dephasing noise acting independently on each qubit with strength $\kappa$.
The dynamics of the state $\varrho_t$ is described by the following Lindblad master equation
\begin{align}
\frac{d\varrho_t}{dt} & = \mathcal{L}_{\omega,\kappa} \varrho_t =  -i \left[ H_\omega,\varrho_t \right] + \frac\kappa{2} \sum_{j=1}^N \mathcal{D} \left[ \sigma_z^{(j)} \right] \varrho_t \notag \\
& =  -i \frac{\omega}{2} \sum_{j=1}^N \left[ \sigma_z^{(j)},\varrho_t \right] + \frac\kappa{2} \left( \sum_{j=1}^N \sigma_z^{(j)} \varrho_t \sigma_z^{(j)} - N \varrho_t \right) \;,
\label{eq:MarkovFreq}
\end{align}
where $\sigma_{z}^{(i)}$ denotes the $z$-Pauli matrix acting on the $i$-th qubit, i.e.
\begin{equation}
\label{eq:sigalpha}
\sigma_z^{(i)} = \underbrace{\id \otimes \dots \otimes \id }_{i-1}\otimes \sigma_z \otimes \underbrace{\id \otimes \dots \otimes \id}_{N-i} \;;
\end{equation}
we have also introduced the dissipation superoperator $\mathcal{D}$, defined as
\begin{equation}
\label{eq:Dsuperop_def}
\mathcal{D}[ A] \bullet =  A \bullet  A^\dag - \frac{1}{2} \left\{ A^\dag  A , \bullet \right\} \;
\end{equation}
and the \emph{Lindbladian} superoperator $\mathcal{L}_{\omega,\kappa}$, the generator of the dynamical semigroup~\cite{Gorini1976,Lindblad1976}.
The Hamiltonian is $H_\omega = \omega J_z$ (we set $\hbar = 1$), where we defined the collective spin operator $J_z = \frac{1}{2} \sum_{j=1}^N \sigma_z^{(j)}$.

Roughly speaking, this situation can be modelled as $N$ trains of independent incoming input bosonic modes, all in the vacuum state, each interacting with one of the qubits for an infinitesimal amount of time.
When the output modes after the interaction are traced away, we get the unconditional evolution described by Eq.~\eqref{eq:MarkovFreq}; for a detailed treatment of this derivation of the Lindblad master equation see, e.g.,~\cite{Ciccarello2017}.

In this work we consider the possibility of continuously measuring the environment causing the dephasing, i.e. sequentially measuring the output modes after the interaction with the system~\cite{Jacobs2006,Genoni2016,Gross2017}.
This corresponds to a (weak) continuous monitoring of the $N$ local observables $\sigma_z^{(j)}$~\cite{wiseman2010quantum,Jacobs2014a}.
Depending on the type of measurement performed on the output modes, we get different \emph{unravellings} of the Markovian master equation~\eqref{eq:MarkovFreq}.
In this work, we will consider photo-detection (PD) and homodyne detection (HD).

The dynamics of the conditional state (denoted as $\varrho^{\text{(c)}}$) of a continuously monitored quantum system is described by stochastic master equations (SMEs)~\cite{Carmichael1993,wiseman2010quantum,Jacobs2014a}.
For time-continuous PD of each dephasing channel (with the same efficiency $\eta$), the evolution is described by the stochastic master equation
\begin{align}
\label{eq:photoSMEparallel}
d\varrho^{\text{(c)}}_t = {} & - i \omega \left[ J_z, \varrho^{\text{(c)}}_t \right] dt + (1-\eta) \frac{\kappa}{2} \sum_j  \mathcal{D} \left[ \sigma_z^{(j)} \right] \varrho^{\text{(c)}}_t \,dt \notag \\
& + \sum_j \left( \sigma_z^{(j)} \varrho^{\text{(c)}}_t \sigma_z^{(j)}- \varrho^{\text{(c)}}_t \right)dN_j  \,,
\end{align}
where the terms $dN_j$ denote independent Poisson increments taking value $0$ (no-click event) or $1$ (detector click event), all with the same average value $\mathbbm{E}\left[ d N_j \right] = \eta \frac{\kappa}{2} dt $.
Because of the unitarity of the collapse operators $\sigma_z$, the statistics of the Poisson processes is independent of the state of the system.
This means that the measurement records $N_j(t)$ contain ``only noise'' and no information about the state and the dynamics.
% , i.e. $\mathcal{F}\left[ p_\text{traj} \right] = 0$.
We notice that, differently from the general case of non-unitary collapse operators, the trace-preserving SME~\eqref{eq:photoSMEparallel} is \emph{linear} in $\varrho_t$.

For time-continuous HD on each output we have a diffusive SME:
\begin{equation}
\label{eq:homoSME}
d\varrho^{\text{(c)}}_t = - i [H_{\omega} , \varrho^{\text{(c)}}_t]\,dt + \frac{\kappa}{2} \sum_j \mathcal{D}[\sigma_z^{j}] \varrho^{\text{(c)}}_t\,dt +  \sqrt{\frac{\eta \kappa}{2}} \sum_j \mathcal{H}[\sigma_z^{(j)} e^{i \theta}]\varrho^{\text{(c)}}_t \, dw_j \:,
\end{equation}
where  $\mathcal{H}[c]\varrho = c \varrho + \varrho c^\dag - \Tr[\varrho (c +c^\dag)]\varrho$.
This is an It\^o stochastic differential equation, where $dw_j = dy_j - 2  \cos \theta \sqrt{\eta} \Tr[\varrho^{\text{(c)}}_t \sigma_z^{(j)}]$ represent independent Wiener increments, formally defined by the identities $dw_j dw_k = \delta_{jk} dt$~\cite{Jacobs2010a}.
The actual measurement results, i.e. the observed photocurrents, are encoded in the stochastic processes $dy_j=dw_j + 2  \cos \theta \sqrt{\eta} \Tr[\varrho^{\text{(c)}}_t \sigma_z^{(j)}]$.
The parameter $\theta$ represents the angle of the HD on the output modes.
Both SMEs~\eqref{eq:homoSME} and~\eqref{eq:photoSMEparallel} give the unconditional Markovian master equation~\eqref{eq:MarkovFreq} when averaged over all the stochastic processes.

While in general the SME~\eqref{eq:homoSME} is non-linear in $\varrho^{\text{(c)}}$, we notice that for the homodyne angle $\theta=\pi/2$ we are again in the situation of observing only noise and getting no information about the evolution, since we get $dy_j=dw_j$.
Similarly to the PD case the SME then becomes linear:
\begin{equation}
\label{eq:homoSMEonlynoise}
d \varrho^{\text{(c)}}_t =  \left( -i \omega \left[ \hat{J}_z ,\varrho^{\text{(c)}}_t \right] + \frac{\kappa}{2} \sum_j \mathcal{D}\left[ \sigma_z^{(j)}  \right] \varrho^{\text{(c)}}_t\right) dt +  i \sqrt{\frac{\eta \kappa}{2} } \sum_j \left[ \sigma_z^{(j)} , \varrho^{\text{(c)}}_t \right] d w_j \;.
\end{equation}

Even if there are cases, as we have shown, in which the trace-preserving SME itself is linear, in general, for any trace-preserving nonlinear SME there is a corresponding linear trace-decreasing SME.
These linear equations are used to obtain closed form solutions~\cite{Jacobs1998,wiseman2010quantum,Jacobs2014a} and they are the fundamental tool to perform statistical inference with continuously monitored systems~\cite{Mabuchi1996}.
The crucial observation is that the norm of the unnormalized states represents a likelihood function for the observed measurement record~\cite{Gammelmark2013a} (even if it does not represent the true probability of the observed photo-current in general~\cite{Wiseman1996}).

\section{Quantum estimation with continuously monitored systems} \label{s:estcontinuous}

\subsection{Quantum estimation theory}
The problem of estimating the unknown value of a real parameter characterizing a probability distribution is a fundamental one in science.
In particular, this statistical setting becomes even more fundamental when dealing with quantum-mechanical systems, since a probabilistic description of an experiment is intrinsically built into the theory.
Quantum estimation theory~\cite{helstrom1976quantum,Holevo2011b,Hayashi2005,Paris2009,Hayashi2017c} studies the problem of how to best estimate parameters encoded in quantum states.

Fundamental results in this field are the classical and quantum Cramér-Rao bounds (CRBs).
The precision of an unbiased estimator, quantified by its standard deviation, is lower bounded as
\begin{align}
\delta \omega \geq \frac{1}{\sqrt{M \mathcal{F}[p(x|\omega)]}} \geq \frac{1}{\sqrt{M \qfi[\varrho_\omega]}}\,, \label{eq:QCRB}
\end{align}
where $M$ is the number of repetitions of the experiment.
The first inequality is the classical CRB for a particular measurement (a positive-operator valued measurement, POVM, with elements $\Pi_x \geq 0 \quad \sum_x \Pi_x = \id$), where $\mathcal{F}[p(x|\omega)] = \mathbbm{E}_p[(\partial_\omega \ln p(x|\omega) )^2]$ is the Fisher information (FI) of the probability distribution, obtained from the Born rule $p(x|\omega)=\mathrm{Tr} \left[ \varrho_\omega \Pi_x \right]$.

By optimizing the FI over all possible POVMs we get the second inequality, the quantum CRB, where the quantum Fisher information (QFI) is defined as~\cite{Braunstein1994}
\begin{equation}
\label{eq:QFIfidelity}
\qfi[\varrho_\omega] = \max_{\text{POVMs}} \mathcal{F}[p(x|\omega)]  =
%4 \frac{D_B \left[ \varrho_\omega, \varrho_{\omega + d \omega} \right] }{d \omega^2} = \frac{ 8 \left(1 - F \left[ \varrho_\omega, \varrho_{\omega + d \omega}  \right] \right) }{ d \omega^2 }
\lim_{\epsilon \rightarrow 0} \frac{ 8 \left(1 - F \left[ \varrho_\omega, \varrho_{\omega + \epsilon}  \right] \right) }{ \epsilon^2 } \,
\end{equation}
in terms of the fidelity $F \left[ \varrho , \sigma  \right] = \left\Vert \sqrt{\varrho} \sqrt{\sigma} \right\Vert_1$, where $\Vert A \Vert_1 = \mathrm{Tr} \left[ \sqrt{A A^\dag} \right]$ is the trace norm.

The quantum CRB can be saturated by performing the optimal local measurement, even though in general the saturation happens asymptotically for $M \to \infty$ using an adaptive procedure~\cite{Barndorff-Nielsen2000}.
From~\eqref{eq:QFIfidelity} we can immediately notice that the QFI is invariant if a parameter-independent unitary is applied to the state $\varrho_{\omega}$, since the fidelity is invariant if the same unitary is applied to both states.
In other words, for any parameter-independent unitary applied to the state, one can apply the inverse unitary to the elements of the original optimal POVM to get the new optimal POVM, which gives exactly the same optimal probability distribution.

\subsection{Ultimate QFI}
In the context of continuously-monitored quantum systems we need to introduce a slightly more general set of tools.
In this scenario, the ultimate limit to the precision in estimating the parameter $\omega$ comes from considering the information contained in the joint state of system and environment $| \Psi_\mathsf{SE} (\omega) \rangle$ (assuming an initial pure state for the system).
We call the QFI of this global state the \emph{ultimate} QFI and we denote it with a bar.
This quantity can be computed from the overlap between the states for different values of the parameter
\begin{align}
\label{eq:ultimQFI}
	\mathcal{\overline{Q}}_{\mathcal{L}_{\omega,\kappa}} = 4 \partial_{\omega_1} \partial_{\omega_2} \log  \left| \langle \Psi_\mathsf{SE}  (\omega_1) |  \Psi_\mathsf{SE}  (\omega_2) \rangle\right| \, \Bigr\vert_{\omega_1 = \omega_2 = \omega}\,\, ,
\end{align}
this expression is equivalent to the general formula~\eqref{eq:QFIfidelity} for pure states.
In this notation we make the dependence on the Lindbladian explicit, to stress that the ultimate QFI does not depend on the specific unravelling.

In general, obtaining an expression for the state $| \Psi_\mathsf{SE} \rangle$ is a hard task, but there is a method based on a modified master equation to compute the fidelity $\langle \Psi_\mathsf{SE}  (\omega_1) |  \Psi_\mathsf{SE}  (\omega_2) \rangle$ without obtaining the state~\cite{Guta2011,Gammelmark2014,Macieszczak2016}.
This technique is fundamental to obtain the results for frequency estimation with continuous monitoring of observables transversal to the Hamiltonian~\cite{Albarelli2018a}.
However, in this paper we deal with collapse operators that commute with the Hamiltonian and calculations are greatly simplified.
As a matter of fact, whenever the collapse operators commute with the Hamiltonian $\hat{H}_\omega$, the ultimate QFI $\overline{\qfi}_{\mathcal{L}_{\omega,\kappa}}$ is equal to the QFI of the state evolving under the unitary dynamics generated by $\hat{H}_\omega$, as shown in~\cite{Albarelli2018a}.

\subsection{Effective QFI}
When a continuous measurement is performed, each observed time-continuous outcome, or trajectory, has a certain probability that we heuristically denote $p_\mathsf{traj}$, independently of the nature of the measurement (PD or HD).
Clearly, if such a probability depends on the parameter $\omega$, it is possible to use this outcomes to learn the value of the parameter and the corresponding CRB depends on the classical FI
\begin{equation}
\label{eq:def_classFI}
\mathcal{F}[p_\mathsf{traj}]=\sum_{\mathsf{traj}} \frac{ (\partial_\omega p_\mathsf{traj} )^2 }{p_\mathsf{traj}} \,,
\end{equation}
where the summation is over all the possible trajectories that can followed by the system.

This quantity can be computed by simulating the stochastic master equation, together with an additional master equation as shown in~\cite{Gammelmark2013a}, see also~\cite{Albarelli2018a} for a stable implementation of this method that takes advantage of the Kraus operator form of the infinitesimal evolution introduced in~\cite{Rouchon2015}.
We also mention that for Gaussian systems, Gaussian measurements and Gaussian dynamics, the computation can be performed more effectively in the  phase-space picture~\cite{Genoni2017}.

The conditional evolution induced by the continuous measurement also allows the possibility to perform a traditional strong measurement on the final conditional state and therefore we have to take into account the QFI of the conditional states.
The correct figure of merit for this setting was introduced in~\cite{Albarelli2017a} and dubbed effective QFI (see also~\cite{Catana2014,Combes2014,Zhang2015f,Shitara2016} where analogous quantities are studied in different contexts).
The effective QFI is defined as the classical FI for the continuous measurement plus the average QFI of the conditional states:
\begin{align}
\widetilde{\qfi}_{\mathsf{unr},\eta}  = \mathcal{F}[p_\mathsf{traj}] + \sum_\mathsf{traj} p_\mathsf{traj} \qfi[\varrho^{\text{(c)}}] \,, \label{eq:effQFIgeneric}
\end{align}
where the subscript $\mathsf{unr}$ stresses that this quantity depends on the particular unravelling, i.e., the particular detection scheme.
This quantity is always greater than or equal to the QFI of the unconditional state (a property called extended convexity of the QFI~\cite{Alipour2015,Ng2016}) and smaller than the ultimate QFI:
\begin{align}\label{eq:QFIineq}
\qfi[\varrho_\mathsf{unc}] &\leq \qunr{\eta}  \leq \overline{\qfi}_{\mathcal{L}_{\omega,\kappa}} \,.
	\end{align}

\section{Results}\label{s:results}
%\subsection{QFI of conditional states}
In this section we show that the effective QFI $\tilde{\qfi}_{\mathsf{unr},\eta}$ for $\eta=1$ can be equal to the QFI of the noiseless unitary dynamics, thus saturating the ultimate bound $\overline{\qfi}_{\mathcal{L}_{\omega,\kappa}}$.
On the other hand,  for $\eta < 1$ the result is the same as for the unconditional dynamics, but with a rescaled coupling constant $\kappa (1-\eta)$.

In particular, we find that this happens when all the information about the parameter is in the conditional states and no information is gained from the continuous monitoring, namely $\mathcal{F}[p_{\sf traj}]=0$ and thus $\widetilde{\qfi}_{\mathsf{unr},\eta} = \sum_\mathsf{traj} p_\mathsf{traj} \qfi[\varrho^{\text{(c)}}]$.

Interestingly, this way of monitoring the environment, where only noise is measured and no information about the state of the system is gained, can be used to reduce decoherence by a factor $(1-\eta)$ by implementing a Markovian feedback~\cite{Szigeti2014}.
We stress that our approach is different because it is not necessary to apply any feedback to take advantage of the improved metrological power of the conditional states.
On the other hand, this strategy works only for parallel noise, while the feedback scheme works for any Hermitian collapse operator, such as transversal noise with collapse operators $\sigma_x^{(j)}$.

In this scenario of ``purely noisy'' monitoring, our result follows from the fact that is possible to recast the conditional evolution as a random unitary transformation followed by the unconditional evolution (the map obtained by exponentiating the Lindbladian) with a rescaled coupling strength $\kappa$.

% The main point is that all the information is contained in the conditional states.

\subsection{Photo-detection}

It is convenient to rewrite the linear PD-SME~\eqref{eq:photoSMEparallel} as\footnote{For brevity, in this section we omit the superscript (c) on the conditional states.}
\begin{align}
\label{eq:photoSMEparallelSuper}
d\varrho_t &=
% \mathcal{H}_{\omega} \varrho_t dt + \mathcal{D}_{(1-\eta)\kappa} \varrho_t dt + \sum_j \left( \mathcal{K}_j -\mathcal{I} \right) \varrho_t dN_j  \, =
\mathcal{L}_{\omega,(1-\eta)\kappa} \varrho_t dt + \sum_j \left( \mathcal{K}_j -\mathcal{I} \right)  dN_j \varrho_t \, ;
\end{align}
where we have introduced the super-operators $\mathcal{K}_j \bullet = \sigma_z^{(j)} \bullet \sigma_z^{(j)} $ and the identity superoperator $\mathcal{I} \bullet = \bullet$.

	We can now exploit the following identity for Poisson processes~\cite[Eq.~(3.162)~p.131]{Jacobs2014a}:
\begin{equation}
\label{eq:PoissonIdentity}
1 + A dN = e^{\log(1+A) dN} = (A +1)^{dN} \;,
\end{equation}
the first equality comes from
\begin{equation}
\begin{split}
e^{\log(1+A) dN} &= \sum_{n=0}^{\infty} \frac{\left[ \log(1+A) dN \right]^n}{n!} = 1 + \sum_{n=1}^{\infty} \frac{\log(1+A)^n dN^n}{n!} = \\
& = 1 + \left(\sum_{n=1}^{\infty} \frac{\log(1+A)^n}{n!} \right) dN = 1 + \left(e^{\log(1+A)}-1\right) dN = \\
& =  1 + A dN \,
\end{split}
\end{equation}
where we used the property of Poisson processes $dN= dN^2 = dN^n$.
The second equality comes straight-forwardly as \(e^{\log(1+A) dN} = e^{\log\left[ (1+A)^{dN}\right] } = (1+A)^{dN}\).

The identity~\eqref{eq:PoissonIdentity} holds not only for scalars but also when $A$ is an operator or a superoperator, the only crucial property is that the exponential is defined as a power series (and in this case $1$ represents the identity operator/superoperator).
In particular we will need the following identities
\begin{equation}
\label{eq:PoissonIdentitySuper}
\mathcal{I} + \left( \mathcal{K}_j - \mathcal{I} \right) dN_j = \mathcal{K}_j^{dN_j} \,,
\end{equation}
that can be obtained in the same way from the following substitutions $1 \mapsto \mathcal{I} $ and $A \mapsto \mathcal{K}_j - \mathcal{I} $.

Using Eq.~\eqref{eq:PoissonIdentitySuper} we can now show that at first order in $dt$ we have \cite{Albarelli2018e}
\begin{equation}
\varrho_{t+dt} = \varrho_t + d \varrho_t = e^{\mathcal{L}_{\omega,(1-\eta)\kappa} dt } \left( \prod_{j=1}^N \mathcal{K}_j^{dN_j} \right) \varrho_t \;.
\end{equation}
We show this by expanding the rhs at first order in $dt$:
\begin{align}
e^{\mathcal{L}_{\omega,(1-\eta)\kappa} dt } \left( \prod_{j=1}^N \mathcal{K}_j^{dN_j} \right) \varrho_t &= \prod_{j=1}^N \left(\mathcal{I} + \mathcal{L}_{\omega,(1-\eta)\kappa} dt  \right) \left[ \mathcal{I} + \left( \mathcal{K}_j - \mathcal{I} \right) dN_j \right] \varrho_t = \label{eq:Expdt1} \\
&= \varrho_t + \mathcal{L}_{\omega,(1-\eta)\kappa} \varrho_t dt + \sum_j^{N}\left( \mathcal{K}_j - \mathcal{I} \right) dN_j \varrho_t = \label{eq:Expdt2}\\
&= \varrho_t + d \varrho_t \label{eq:Expdt3}.
\end{align}
To get from \eqref{eq:Expdt1} to \eqref{eq:Expdt2} we have used the property of Poisson processes $dt \,dN_j =0$ as well as the fact the all the processes are independent, so that $dN_j dN_k = \delta_{jk}\, d N_k$; finally to get to~\eqref{eq:Expdt3} we have used the stochastic increment~\eqref{eq:photoSMEparallelSuper}.
% In Eq.~\eqref{eq:photoSMEparallel} all the terms multiplying the Poisson increments are given by the superoperator $\mathcal{K}_j - \mathcal{I}$ (acting locally on the $j$-th qubit), where $\mathcal{K}_j \bullet = \sigma_z^{(j)} \bullet \sigma_z^{(j)} $ and $\mathcal{I} \bullet = \bullet$ is the identity superoperator.

% By using this observation in conjunction with the identity~\eqref{eq:PoissonIdentity} and by exploiting the properties of independent Poisson increments, $dt dN_j = 0$ and $d N_j dN_k = \delta_{jk} dN_j$,
We can rewrite the infinitesimal evolution of the density operator more explicitly as
% \begin{equation}
% \varrho_{t+dt} = e^{dt (1-\eta)\frac{\kappa}{2} \sum_j \mathcal{D}[\sigma_z^{(j)}] \,} \left[ \left( \prod_j {\sigma_z^{(j)}}^{dN_j} \right) \left(  e^{-i \omega \hat{J}_z \, dt} \varrho_{t} \, e^{i \omega \hat{J}_z \, dt} \right)  \left(\prod_j {\sigma_z^{(j)}}^{dN_j}\right) \right] \;,
% \end{equation}
\begin{equation}
\varrho_{t+dt} =  \left( \prod_j {\sigma_z^{(j)}}^{dN_j} \right) \left( e^{\mathcal{L}_{\omega,(1-\eta)\kappa} dt\,}   \varrho_{t} \, \right) \left(\prod_j {\sigma_z^{(j)}}^{dN_j}\right) \;,
\end{equation}
where we have also exchanged the order of the action of the superoperators, since they commute.
This identity is true up to order $dt$ and we can think about it as the unconditional dynamics with a rescaled coupling $\kappa (1-\eta)$, followed by random ``spin-flips''.

Since all the superoperators applied to the state commute, the solution is trivial and iterating the infinitesimal evolution amounts to integrating the various exponents separately, therefore we get to
% \begin{equation}
% \label{eq:ParallelPoissonEvo}
% \begin{split}
% \varrho_{t} &=  e^{(1-\eta) \frac{\kappa}{2} \sum_j \mathcal{D} \left[\sigma_z^{(j)} \right] \, t} \left[ \left( \prod_j {\sigma_z^{(j)}}^{ N_j(t)} \right) \left(  e^{-i \omega \hat{J}_z \, t} \varrho_{0} \, e^{i \omega \hat{J}_z \, t} \right)  \left(\prod_j {\sigma_z^{(j)}}^{N_j(t)}\right) \right]\\
%  &=  e^{(1-\eta) \frac{\kappa}{2} \sum_j \mathcal{D} \left[\sigma_z^{(j)} \right] \, t} \left[ e^{-i \frac{\pi}{2}\sum_j \sigma_z^{(j)} N_j(t)} \left(  e^{-i \omega \hat{J}_z \, t} \varrho_{0} \, e^{i \omega \hat{J}_z \, t} \right)  e^{i \frac{\pi}{2}\sum_j \sigma_z^{(j)} N_j(t)} \right] \;,
% \end{split}
% \end{equation}
\begin{equation}
\label{eq:ParallelPoissonEvo}
\begin{split}
\varrho_{t} &=  \left( \prod_j {\sigma_z^{(j)}}^{ N_j(t)} \right) \left( e^{ \mathcal{L}_{\omega,(1-\eta)\kappa} t \,}   \varrho_{t} \,  \right) \left(\prod_j {\sigma_z^{(j)}}^{N_j(t)}\right) \\
&=  e^{-i \frac{\pi}{2}\sum_j \sigma_z^{(j)} N_j(t)} \left(  e^{ \mathcal{L}_{\omega,(1-\eta)\kappa} t \,}   \varrho_{t} \,   \right)  e^{i \frac{\pi}{2}\sum_j \sigma_z^{(j)} N_j(t)} \;,
\end{split}
\end{equation}
where the random variable $N_j(t) = \int_{0}^t dN_j$ counts the number of detections at the $j$-th detector.

Since all the operators appearing in this formula are commuting, the evolution of each conditional state can also be obtained as the solution of the master equation \eqref{eq:MarkovFreq} with a reduced dephasing rate $(1-\eta)\kappa$ and an extra term in the Hamiltonian:
\begin{align}
H = H_\omega + \frac{\pi}{2} \sum_j \beta_j(t) \,\sigma_z^{(j)}\,,
\end{align}
where $\beta_j(t) = dN_j / dt$ is a stochastic term, satisfying the Poissonian statistics defined above.
By averaging over the stochastic process, one obtains the unconditional dephasing dynamics, while a continuous observation in time of the classical stochastic process allows to obtain the conditional states.
It is in fact well known that Markovian dephasing is equivalent to the average dynamics obtained by putting a stochastic term in the system Hamiltonian~\cite{Crow2014,Rossi2015,Burgarth2016,Chen2019c,Fagnola2019}.
Remarkably, we have derived the same results starting from a fully quantum approach, that is, considering the interaction of the system with a quantum environment, represented by trains of incoming bosonic \emph{input} modes, with the classical stochastic terms appearing because of the measurement performed on the environment.

The spin flips depending on the random Poisson processes are parameter-independent unitaries and thus they do not affect the QFI of the state.
It follows that the QFI of each conditional state is exactly equal to the QFI of the unconditional one, but with a rescaled coupling $\kappa \mapsto \kappa (1-\eta)$.
For $\eta = 1$ the QFI of each conditional state (and thus also the average QFI) is equal to the noiseless case and saturates the ultimate bound $\overline{\qfi}_{\mathcal{L}_\omega}$.

The fact that for perfectly efficient detectors is possible to re-obtain the noiseless QFI was already shown in~\cite{Albarelli2018a} for GHZ initial states by means of a more intuitive argument.
On the other hand, the effect of inefficient detection was only inferred from the numerical simulations, while we now have an analytical argument, that also applies for any input state of $N$ qubits.

\subsection{Homodyne detection}

In our previous paper~\cite{Albarelli2018a} we considered HD with angle $\theta=0$ and we have observed numerically that this unravelling is not optimal, since it does not give an effective QFI equal to the noiseless case, even for perfectly efficient detection.
Here, we show that by measuring the appropriate quadrature it is possible to leave all the information in the conditional states, while getting no information from the observed photocurrent, exactly as in the PD case.

As shown previously, by choosing $\theta = \pi / 2$ the observed photocurrent is again only noise and no signal, i.e. $d y_t = d w_t$ and the corresponding classical FI vanishes.
With this choice the dynamics is described by the SME~\eqref{eq:homoSMEonlynoise}, that we can rewrite compactly as
\begin{equation}
\label{eq:homoSMEparallelSuper}
d\varrho_t =
\mathcal{L}_{\omega,\kappa} \varrho_t dt +  i \sqrt{\frac{\eta \kappa}{2} } \sum_j \left[ \sigma_z^{(j)} , \varrho_t \right] d w_j \,,
\end{equation}
Similarly to the case of Poissonian processes, we can write the infinitesimal evolution
\begin{equation}
\varrho_{t+dt} = e^{ \mathcal{L}_{\omega,(1-\eta)\kappa} dt}
\left(   e^{i \sqrt{\frac{\eta \kappa}{2}} \sum_j dw_j \sigma_z^{(j)}} \varrho_{t} \,  e^{-i \sqrt{\frac{\eta \kappa}{2}} \sum_j dw_j \sigma_z^{(j)}}  \right) \,.
\end{equation}
This identity can be verified by expanding the exponentials and keeping the terms up to order $dt$, keeping in mind the properties of the Wiener increments $d w_j d w_k = \delta_{jk} dt$.
As in the previous case, the equation can be integrated easily, since all the operators commute with each other, and we can rearrange the terms to rewrite the evolved conditional state as
% \begin{equation}
% \varrho_{t} = e^{(1-\eta) \frac{\kappa}{2} \sum_j \mathcal{D} \left[\sigma_z^{(j)} \right] \, t} \left[  e^{i \sqrt{\frac{\eta \kappa}{2}} \sum_j W_j (t) \sigma_z^{(j)}} \left(  e^{-i \omega \hat{J}_z \, t} \varrho_{0} \, e^{i \omega \hat{J}_z \, t} \right) e^{-i \sqrt{\frac{\eta \kappa}{2}} \sum_j W_j (t) \sigma_z^{(j)}}  \right] \;,
% \end{equation}
\begin{equation}
\varrho_{t} = e^{i \sqrt{\frac{\eta \kappa}{2}} \sum_j W_j (t) \sigma_z^{(j)}} \left( e^{\mathcal{L}_{\omega,(1-\eta)\kappa} t } \varrho_{0} \,  \right) e^{-i \sqrt{\frac{\eta \kappa}{2}} \sum_j W_j (t) \sigma_z^{(j)}}  \;,
\end{equation}
where the random variables $W_j (t) = \int_0^t d w_j$ are normally distributed with mean $0$ and variance $t$.
% and therefore $\sum_j W_j (t) $ is a random variable with mean $0$ and variance $N t$.
As in the case of continuous photo-detection, the formula above shows that the evolution of the conditional states is completely equivalent to the one that we would obtain by considering the Markovian master equation \eqref{eq:MarkovFreq} with a reduced dephasing rate $(1-\eta)\kappa$ and then by modifying the Hamiltonian as
\begin{align}
H = H_\omega - \sqrt{\frac{\eta\kappa}{2}} \sum_j \xi_j(t) \, \sigma_z^{(j)} \,,
\end{align}
where $\xi_j(t)=dw_j/dt$ is a \emph{white noise} stochastic term.

We can see again that each conditional state differs from a purely Hamiltonian evolution only by a random unitary transformation.
For $\eta=1$, the QFI of each conditional state (and thus also the average QFI) is equal to the noiseless case and therefore we can saturate the ultimate bound even with homodyne monitoring. On the other hand, for inefficient monitoring, one obtains the same results of noisy case, but with a reduced effective dephasing rate $(1-\eta)\kappa$.
\section{Conclusions}\label{s:conclusions}
We have shown that by continuously monitoring the dephasing channels acting on $N$-qubit systems it is possible to recast the conditional evolution as a random unitary (which roughly speaking depends on the sum of all the observed outcomes) plus the unconditional dephasing evolution with a decreased strength $(1-\eta) \kappa$.

From the point of view of quantum metrology, this results implies that the QFI of the conditional states is equal to the QFI of the unconditional dynamics with a rescaled dephasing strength, since the random unitaries do not depend on the frequency and thus do not affect the QFI.
For perfect efficiency, the QFI of all the conditional states is equal to the QFI of the initial state evolved noiselessly.
This result is in agreement with the more general results obtained in~\cite{Gregoratti2003}, where it is shown how it is in general possible to perfectly counteract the effect of a noisy channel by measuring the environment \emph{if and only if} the open evolution can be written as an ensembles of random unitaries and thus the measurement on the environment gives no information about the input state.
While the perfect cancellation of the noisy channel explained in~\cite{Gregoratti2003} requires in general a final unitary dependent on the measurement outcomes, we have not considered this step, since the QFI is invariant under unitary transformations.
However, we remark that in general the observed outcomes will be needed to implement the optimal measurement.

We stress that, differently from our previous work~\cite{Albarelli2018a}, these new results are independent from the initial state.
If we restrict to a GHZ initial state, the relevant quantities that describe the conditional states for photo-detection and homodyne are respectively $\bar{N}(t) = \sum_j N_j(t)$ and $\bar{W}(t) =\sum_j W_j(t)$.
From a physical and practical point of view, this shows that in this case one does not need to monitor the $N$ independent channels with $N$ detectors, since the same conditional evolution can be obtained by employing a single detector that monitors the corresponding total photo-counts or photo-current.

We finally remark that these results can be readily extended to the case of non-linear quantum metrology in many-body open quantum systems undergoing single-body and many-body dephasing~\cite{Beau2017}.

\section*{Acknowledgments}
F.A. acknowledges support from the UK National Quantum Technologies Programme (EP/M013243/1).
M.A.C.R. acknowledges financial support from the Academy of Finland via the Centre of Excellence program (Projects no.~312058 and no,~287750).
M.G.G. acknowledges support from a Rita Levi-Montalcini fellowship of MIUR.
\bibliographystyle{apsrev4-2}
\bibliography{parallel-opt}

\end{document}